\documentstyle[epsf]{tmu-vpi}
\begin{document}

\title{%
Constraining New Physics with Vertex Corrections
}

\author{%
Tatsu Takeuchi${}^{(1)}$, Oleg Lebedev${}^{(1)}$, and 
Will Loinaz${}^{(1,2)}$\\
{\it ${}^{(1)}$Department of Physics, Virginia Tech, Blacksburg, VA 24061, U.S.A.}\\
{\it ${}^{(2)}$Department of Physics, Amherst College, Amherst MA 01002, U.S.A.}\\
{\it takeuchi@vt.edu, lebedev@vt.edu, loinaz@alumni.princeton.edu}
}

\maketitle

\section*{Abstract}

We discuss how new physics not encompassed within the $STU$ oblique 
correction framework can be constrained from precision
electroweak measurements via vertex corrections to $Z$--pole observables.

\section{Limitations of the Oblique Correction Analysis}

In constructing models beyond the Standard Model (SM) such as technicolor or
supersymmetric theories, it is important to test their viability by looking 
for possible conflicts with currently available experimental data.  
A popular procedure has been the $STU$ oblique correction analysis of
Ref.~\cite{STU}.  However, the method can only be applied to theories
which satisfy three conditions, namely (1) the electroweak gauge group is
the standard $SU(2)_L \times U(1)_Y$, (2) vertex and box corrections from
new physics are negligible, and (3) the scale of new physics is large
compared to the electroweak scale.   Furthermore, the oblique correction 
analysis requires (4) the model to be completely specified since any particle
with electroweak gauge quantum numbers will show up in a vacuum polarization
loop.  Many theories of interest do not
fall into this category.  For instance, in the topcolor assisted technicolor
model of Ref.~\cite{Hill}, (1) the electroweak gauge group
is $SU(2)_L \times U(1) \times U(1)$ resulting in an extra $Z'$ boson, 
(2) there are potentially large corrections to the $Zf\bar{f}$ vertex for 
third generation fermions coming from $Z'$ and coloron exchange, 
(3) the scale of new physics may be as low as a TeV, and
(4) the technisector of the theory remains unspecified.

\section{Vertex Corrections at the $Z$--pole}

The solution to our predicament lies in the observation that a large number
of electroweak observables have been measured accurately at LEP and SLD 
{\it on the $Z$--pole}, and that the majority of them are just 
{\it ratios of coupling constants} of the $Z$ to quarks and leptons.   
This means that they are
only sensitive to oblique corrections through the effective value of
$\sin^2\theta_W$ {\it at that particular energy scale}. 
Therefore, one can use just one of the observables,
$A_{\rm LR}$ say, to {\it fix} the value of $\sin^2\theta_W$, and use it
to predict all the rest for the SM.  
Any deviations of the data from those predictions must be due to 
{\it vertex corrections} from new physics.

This method of comparing SM predictions to $Z$--pole 
electroweak observables solves
all of the problems mentioned above since (1) mixing of the $Z$ with extra
gauge bosons can be treated as a vertex correction, (2) vertex corrections
need not be small while box corrections are still naturally suppressed
on the $Z$--pole, (3) the new physics scale need not be large since
only observables at one scale are used in the analysis, and (4) the
theory need not be specified completely since the method is blind
to oblique corrections.

\section{Applications}

This method has been used successfully to 
constrain the size of possible vertex corrections 
to the $Zb\bar{b}$ vertex \cite{Zbb}, 
and to constrain the topcolor assisted technicolor 
model mentioned above \cite{topcolor}.
We have also applied the technique to constrain:
an R--parity violating extension to the MSSM \cite{Rparity},
Higgs masses in a general two Higgs doublet model \cite{twohiggs}, 
and the fundamental Planck scale of a model with large extra 
dimensions \cite{torsion}.
Details are provided in the respective references.

\section*{Acknowledgements}

This work was supported in part
by the U.S. Department of Energy, grant DE--FG05--92--ER40709, Task~A.



\begin{thebibliography}{99}
\bibitem{STU}
M.~E.~Peskin and T.~Takeuchi,
Phys. Rev. Lett. 65 (1990) 964;
Phys. Rev. D46 (1992) 381.

\bibitem{Hill}
C. T. Hill, Phys. Lett. B345 (1995) 483.

\bibitem{Zbb}
T.~Takeuchi, A.~K.~Grant, and J.~L.~Rosner, 
in the {\it Proceedings of DPF'94},
[hep-ph/9409211].

\bibitem{topcolor}
W.~Loinaz and T.~Takeuchi,
Phys. Rev. D60 (1999) 015005.

\bibitem{Rparity}
O.~Lebedev, W.~Loinaz, and T.~Takeuchi,
Phys. Rev. D61 (2000) 115005;
Phys. Rev. D62 (2000) 015003.

\bibitem{twohiggs}
O.~Lebedev, W.~Loinaz, and T.~Takeuchi,
hep-ph/0002106.

\bibitem{torsion}
L.~N.~Chang, O.~Lebedev, W.~Loinaz, and T.~Takeuchi,
hep-ph/0005236.

\end{thebibliography}
\end{document}